\documentclass[
    twocolumn,
	prd,
	amssymb,
	preprintnumbers,superscriptaddress,
	nofootinbib]{revtex4-1}

\pdfoutput=1

\usepackage{graphicx}
\usepackage{enumitem}
\usepackage{latexsym}
\usepackage{amsfonts}
\usepackage{amssymb}
\usepackage{xcolor}
\usepackage[export]{adjustbox}
\usepackage{amsmath}
\usepackage[thinlines]{easytable}
\usepackage{slashed}
\usepackage{dcolumn}
\usepackage{verbatim}
\usepackage{float}
\usepackage{multirow}
\usepackage{xspace}
\usepackage[normalem]{ulem}
\usepackage[
pdfauthor={Jeremy Sakstein}]{hyperref}
\usepackage{tabularx}
\usepackage{paralist}

\renewcommand{\tilde}{\widetilde} 

\newcommand{\beq}{\begin{equation}}
\newcommand{\eeq}{\end{equation}}
\newcommand{\bea}{\begin{eqnarray}}
\newcommand{\eea}{\end{eqnarray}}

\newcommand{\dd}{\mathrm{d}}
\newcommand{\mpl}{M_{\rm Pl}}

\DeclareRobustCommand{\okina}{%
  \raisebox{\dimexpr\fontcharht\font`A-\height}{%
    \scalebox{0.8}{`}%
  }%
}

\allowdisplaybreaks

\begin{document}

\title{Neutrino-Assisted Early Dark Energy is a Natural Resolution of the Hubble Tension}
\author{Mariana Carrillo Gonz\'alez} 
\email{m.carrillo-gonzalez@imperial.ac.uk}
\affiliation{Theoretical Physics, Blackett Laboratory, Imperial College, London, SW7 2AZ, U.K}
\author{Qiuyue Liang} 
\email{qyliang@sas.upenn.edu}
\affiliation{Center for Particle Cosmology, Department of Physics and Astronomy, University of Pennsylvania, Philadelphia, Pennsylvania 19104, USA }
\author{Jeremy Sakstein} \email{sakstein@hawaii.edu}
\affiliation{Department of Physics \& Astronomy, University of Hawai\okina i, Watanabe Hall, 2505 Correa Road, Honolulu, HI, 96822, USA}
\author{Mark Trodden} 
\email{trodden@upenn.edu}
\affiliation{Center for Particle Cosmology, Department of Physics and Astronomy, University of Pennsylvania, Philadelphia, Pennsylvania 19104, USA }

\date{\today}

\begin{abstract}
It has very recently been claimed~\cite{deSouza:2023sqp} that the neutrino-assisted early dark energy model --- a promising resolution of the Hubble tension that can ameliorate the theoretical fine-tuning and coincidence problems that plague other theories --- does not provide natural or cosmologically interesting results.~In this short paper, we show that these conclusions are incorrect for three reasons.~First, we identify errors in the calculations.~Second, we dispute the definition in~\cite{deSouza:2023sqp} of what constitutes an ``interesting" and ``natural" model.~Finally, we demonstrate that the conclusions of~\cite{deSouza:2023sqp} were arrived at without fully exploring the full parameter space of the model.~Neutrino-assisted early dark energy remains a natural and interesting potential resolution of the Hubble tension that merits further study.
\end{abstract}

\maketitle

\section{Introduction}

The Hubble tension \cite{Verde:2019ivm,Knox:2019rjx,DiValentino:2021izs,Kamionkowski:2022pkx} is one of the biggest mysteries confounding modern cosmologists.~Despite a continued updating and interrogation of various cosmological datasets over the last decade, the disagreement between early- and late-time measurements of the Hubble constant $H_0$  has persisted and worsened to the point where the discrepancy has surpassed $5\sigma$ \cite{Verde:2019ivm,Knox:2019rjx,Riess:2021jrx,Kamionkowski:2022pkx,Brout:2022vxf}.~The inability of the $\Lambda$CDM  cosmological standard model to account for all of the astrophysical and cosmological observations has motivated theorists to consider the tantalizing possibility that the Hubble tension is a signal of new physics beyond the cosmological standard model.~

A plethora of theoretical models that can resolve the tension have been proposed that have met with varying degrees of success \cite{Knox:2019rjx,Schoneberg:2021qvd,DiValentino:2021izs}.~Among the various proposals, Early Dark Energy (EDE) \cite{Poulin:2018cxd} has emerged as a promising candidate \cite{DiValentino:2021izs,Schoneberg:2021qvd}.~In this scenario, a minimally-coupled scalar  field $\phi$ is frozen at some initial condition $\phi_i$ at early times, but begins to roll around the epoch of matter-radiation equality (MRE). During this phase of rolling, the scalar accounts for $\sim10\%$ of the energy budget of the universe and increases the Hubble parameter compared to $\Lambda$CDM.~This has the effect of decreasing the sound horizon, which inversely increases the early-time measurements of $H_0$ so that they are consistent with the (larger) late-time measurements (see \cite{Knox:2019rjx,Schoneberg:2021qvd,Kamionkowski:2022pkx} for a detailed explanation of this mechanism).~

The minimal EDE scenario suffers from theoretical fine-tunings and a coincidence problem.~The scalar field mass must be fine-tuned to $m_\phi\sim 10^{-29}$eV (the Hubble scale at MRE) in order for it to transit from the over- to under-damped regime at this epoch.~Such small scalar masses present a technical-naturalness challenge for EDE models, since  quantum corrections will drive the mass towards the cut-off of the effective field theory (EFT).~Additionally, the physics of MRE is completely disconnected from the physics of the scalar presenting a coincidence (or \textit{why now?}) problem for EDE models.~Why should the onset of EDE occur at MRE and not some other time?

Two of us have proposed a framework that ameliorates a number of the theoretical issues with EDE --- neutrino-assisted early dark energy ($\nu$EDE) \cite{Sakstein:2019fmf}.~Here, EDE has a Yukawa coupling to neutrinos, which are relativistic in the early universe but become non-relativistic when the temperature of the universe is of order their mass.~
When this happens, the neutrinos inject energy into the scalar, giving it a ``kick".~It is a cosmic coincidence that the sum of the neutrino masses is of order the temperature at MRE.~Thus, if the neutrino mass spectrum is dominated by one species, the kick naturally happens at approximately the correct time that EDE needs to become active to resolve the $H_0$ tension without the need to fine-tune the scalar field  mass.~Similarly, there is no need to fine-tune the initial conditions because the scalar can begin at its minimum --- the natural initial condition --- and be displaced by the kick.~In subsequent work~\cite{CarrilloGonzalez:2020oac}, the four of us calculated the quantum corrections to the scalar field mass in this setup and found that a light mass is technically natural, thereby resolving the final EDE fine-tuning.~We also constructed numerical solutions and explored how to generalize the model (including the form of the conformal coupling) in order to ensure that the EFT remains well behaved at high redshifts.~$\nu$EDE is therefore a theoretically appealing potential resolution of the Hubble tension.

In a recent paper~\cite{deSouza:2023sqp}, it has been claimed that there are no regions of the $\nu$EDE parameter space where the scenario is ``natural" or ``cosmologically interesting".~The purpose of this paper is to demonstrate that these claims are incorrect.~The authors of~\cite{deSouza:2023sqp} make three specific claims
\begin{compactenum}
    \item That there is a maximum magnitude of the kick that prevents the success of the mechanism;
    \item That there are no models with the initial condition $\phi_i=0$ (which they claim is the ``natural" one) that can inject the correct amount of EDE at MRE to resolve the tension; and
    \item That there are no ``interesting" models with $\phi_i\ne0$ because the authors consider these initial conditions to be unnatural, and they find  uncoupled models with the same initial conditions that can also provide sufficient EDE to resolve the tension.
\end{compactenum}
In what follows, we will demonstrate why we strongly disagree with these claims.

We will first show that the equations in~\cite{deSouza:2023sqp} used to arrive at claim (1) are incorrect, arising from the use of an unphysical neutrino energy density and pressure. We next explain why claims (2) and (3) would be incorrect. In brief, the natural initial condition is not $\phi_i=0$ because, as we discussed at length in \cite{CarrilloGonzalez:2020oac}, the coupling to neutrinos shifts the minimum of the effective potential governing the scalar field dynamics away from zero and towards large values.~Large initial values of the field, that are dismissed as unnatural in~\cite{deSouza:2023sqp}, are, in fact natural, and the ensuing phenomenology remains interesting.

\section{Neutrino-Assisted Early Dark Energy Is Natural and Interesting}

In this section we take each of the claims above in turn and explain why we disagree with the results in~\cite{deSouza:2023sqp}.

\subsection{Derivation of the $\nu$EDE Equations of Motion}

The action for $\nu$EDE is\footnote{We will not specify the potential $V(\phi)$ in what follows, since our conclusions are general and the $\nu$EDE framework works for a generic potential.~The models studied in \cite{Sakstein:2019fmf,CarrilloGonzalez:2020oac} were chosen to be of the form $V(\phi)\sim\lambda\phi^4$, so that the potential is simple and renormalizable.}
\begin{align}
    S&=\int\dd ^4x\sqrt{-g}\left[-\frac12\partial_\mu\phi\partial^\mu\phi-V(\phi)\right]\nonumber\\&+S_m[g_{\mu\nu};\Psi_m]+S_\nu[\bar{g}_{\mu\nu};\Psi_\nu],\label{eq:action}
\end{align}
where $S_m$ is the action for all matter fields $\Psi_m$, and $S_\nu$ is the action for neutrino fields $\Psi_\nu$.~This implies that all matter fields except neutrinos move on geodesics of the Einstein frame metric, $g_{\mu\nu}$, and neutrinos move on geodesics of the Jordan frame metric, $\bar{g}_{\mu\nu}$, to which they couple minimally.~The two metrics are related via
\begin{equation}
  \bar{g}_{\mu\nu}=A^2(\phi)g_{\mu\nu} \ , \quad \text{with} \quad A(\phi)=\exp(\beta\phi/\mpl) \ .
\end{equation}
The cosmological equation of motion for the scalar resulting from the action \eqref{eq:action} is (assuming a flat FLRW metric)
\begin{equation}
\label{eq:scalar_eom}
\ddot{\phi}+3H\dot\phi+V'(\phi)=\frac{\beta}{\mpl}\Theta(\nu)
\end{equation}
where $\Theta(\nu)=g_{\alpha\beta}\Theta^{\alpha\beta}(\nu)$ is the trace of the Einstein frame energy momentum tensor $\Theta^{\alpha\beta}(\nu)=2/\sqrt{-g}\delta S_\nu/\delta g_{\alpha\beta}$.~This is not covariantly conserved ($\nabla_\alpha\Theta^{\alpha\beta}\ne0$) due to the non-minimal coupling between the scalar and neutrinos.~In contrast, the Jordan frame energy-momentum tensor $\bar{\Theta}^{\alpha\beta}(\nu)=2/\sqrt{-\bar{g}}\delta S_\nu/\delta \bar{g}_{\alpha\beta}$ is covariantly conserved with respect to the Jordan frame connection i.e., $\bar{\nabla}_\alpha\bar{\Theta}^{\alpha\beta}(\nu)=0$.~This implies that we should apply thermodynamics to derive the neutrinos' pressure and density in the Jordan frame and translate all quantities into the Einstein frame.~The two energy-momentum tensors are related by the following formulae \cite{Sakstein:2014jrq}:
\begin{align}
    {\Theta}^{\alpha\beta}(\nu)&=A^6\bar{\Theta}^{\alpha\beta}(\nu),\quad{\Theta}^{\alpha}_{\phantom{\alpha}\beta}(\nu)=A^4\bar{\Theta}^{\alpha}_{\phantom{\alpha}\beta}(\nu),\nonumber\\{\Theta}_{\alpha\beta}(\nu)&=A^2\bar{\Theta}_{\alpha\beta}(\nu),\quad\quad {\Theta}(\nu)=A^4\bar{\Theta}(\nu).\label{eq:relation_between_jordan_and_einstein_frame_EM_tensors}
\end{align}
This implies that the Jordan and Einstein frame pressure and density are related via
\begin{equation}
\label{eq:relation_between_jordan_and_einstein_frame_density_and_pressure}
P_\nu=A^4\bar{P}_\nu,\quad \rho_\nu=A^4\bar{\rho}_\nu.
\end{equation}
It is $\bar{P}_\nu$ and $\bar{\rho}_\nu$ that must be calculated using the Fermi-Dirac distribution.~Doing so, one finds
\begin{align}
    \nonumber\bar{\Theta}(\nu)&=3\bar{P}_\nu-\bar{\rho}=-\frac{g_\nu}{2\pi^2}\bar{T}_\nu^4\tau\left(\frac{{m}_\nu}{\bar{T}_\nu}\right);\\
\tau(x)&=x^2\int_x^\infty\frac{(u^2-x^2)^\frac12}{e^u+1}\dd u,\label{eq:3P-rho_Jordan_frame}
\end{align}
where, since neutrinos decouple while being relativistic, the Jordan frame temperature is $\bar{T}_\nu=\bar{T}_0/\bar{a}$ with $\bar{a}$ the Jordan frame scale factor.~The relation between the Jordan and Einstein frame temperature $T_\nu$ is given by\footnote{We can relate the Jordan and Einstein frame temperatures as follows.~First, we need to relate the two scale factors.~The line-element in the Jordan frame is
\begin{align*}
\dd\bar{s}^2&=\bar{g}_{\mu\nu}\dd x^\mu\dd x^\nu=A^2(\phi)\dd s^2\nonumber\\&=-A^2(t)\dd t^2+A^2(t)a^2(t)\delta_{ij}\dd x^i\dd x^j,\label{eq:ds2_relation_between_Jordan_and_einstein_frames}
\end{align*}
where $\dd s^2=-\dd t^2+a(t)^2\delta_{ij}\dd x^i\dd x^j$ with $a(t)$ the Einstein frame scale factor and $A(t)=A(\phi(t))$.~Defining the Jordan frame coordinate time $\bar{t}(t)$ by $\dd\bar{t}=A(t)\dd t$ the Jordan frame line-element can be brought into the standard coordinate time form:
\[
    \dd\bar{s}^2=-\dd{\bar{t}}^2+A^2(\bar{t})a^2(\bar{t})\delta_{ij}\dd x^i\dd x^j=-\dd{\bar{t}}^2+\bar{a}^2(\bar{t})\delta_{ij}\dd x^i\dd x^j.
\]
From this, one can see that $\bar{a}=A a$ and $\bar{T}_\nu=T_\nu/A$.} $\bar{T}_\nu=T_\nu/A(\phi)$.~Returning to equation \eqref{eq:3P-rho_Jordan_frame}, one has
\begin{equation}
    \bar{\Theta}(\nu)=3\bar{P}_\nu-\bar{\rho}=-\frac{g_\nu }{2\pi^2A^4}{T}_\nu^4\tau\left(\frac{m_\nu}{\bar{T}_\nu}\right) \ .
\end{equation}
Using equation \eqref{eq:relation_between_jordan_and_einstein_frame_EM_tensors}, equation \eqref{eq:scalar_eom} becomes
\begin{equation}
\label{eq:scalar_eom_using_Jordan_frame_trace}
\ddot{\phi}+3H\dot\phi+V'(\phi)=-\frac{g_\nu\beta}{2\pi^2\mpl}T_\nu^4\tau\left(\frac{m_\nu}{\bar{T}_\nu}\right).
\end{equation}
This is identical to the formula derived in \cite{Sakstein:2019fmf}, except for the argument of $\tau$ which involves $\bar{T}_\nu=T_\nu/ A(\phi)$ instead of $T_\nu$, which we used in \cite{Sakstein:2019fmf,CarrilloGonzalez:2020oac} as an approximation valid in the limit $\beta\phi/\mpl\ll1$ in order to ensure that the model constituted a healthy EFT.~In this limit,the temperatures in both frames are equivalent, whereas away from this limit, there is a small shift in the time that the energy is injected.~Importantly the factor of $A(\phi)^4$ that arises when transforming the energy-momentum tensor from the Jordan to the Einstein frame is cancelled when the temperature is similarly transformed.~Independent of the potential or the time of the injection, this equation of motion predicts a kick of order $\Delta\phi\approx-0.03\beta\mpl$ \cite{Sakstein:2019fmf}, as long as we remain within the regime of validity of the EFT.~Note that over this range, this quantity increases linearly with $\beta$.

\subsection{Comparison with the Derivation in~\cite{deSouza:2023sqp}}

In~\cite{deSouza:2023sqp}, the derivation begins from the scalar field equation of motion (our equation \eqref{eq:scalar_eom} and equation (A1) in the appendix of~\cite{deSouza:2023sqp}):
\begin{equation}
\label{eq:Souza_Rosenfeld_scalar_EOM}
\ddot{\phi}+3H\dot\phi+V'(\phi)=\frac{\beta}{\mpl}\Theta(\nu)=\frac{\beta}{\mpl}(3P-\rho).
\end{equation}
Next, 
an equivalent expression is used
\begin{equation}
\label{eq:Souza_Rosenfeld_scalar_EOM_new_P_rho}
\ddot{\phi}+3H\dot\phi+V'(\phi)=\frac{\beta}{\mpl}\tilde{\Theta}(\nu) e^{\beta\frac{\phi}{\mpl}} \ .
\end{equation} 
with $\tilde{P}=P\exp(-\beta\phi/\mpl)$, $\tilde{\rho}=\rho$, and $\tilde{\Theta}(\nu)=(3\tilde{P}-\tilde{\rho})$.~This can be thought of as a simple redefinition of variables, and so is certainly allowed.~However, crucially, the authors of~\cite{deSouza:2023sqp} then set
\begin{equation}
\label{eq:Souza_Rosenfeld_incorrect_relation_of_P_rho_to_tau}
\tilde{\Theta}(\nu)=-\frac{g_\nu}{2\pi^2}{T}_\nu^4\tau\left(\frac{m_\nu}{{T}_\nu}\right)
\end{equation}
(equation (A3) of the appendix). In our view this step is flawed because $\tilde{P}$ and $\tilde{\rho}$ are not a physical pressure and density\footnote{Some authors refer to these quantities as the \textit{conserved} density and pressure \cite{Khoury:2003rn,Brax:2004qh,Sakstein:2014jrq}.~This is because these quantities satisfy the FLRW continuity equation in the Einstein frame, and so redshift as they would if the scalar were uncoupled. This does not however imply that they are physical.~They are quantities that are sometimes useful for calculation purposes.}  The Jordan frame is the unique frame in which the neutrinos obey Fermi-Dirac statistics. Although one can make a redefinition of variables to make the computation easier, the physics should not depend on the choice of frame or thermodynamics variables.~Because of this, the subsequent equations in~\cite{deSouza:2023sqp} differ from the correct equations, and are only approximately in agreement with them in the limit $\beta\phi/\mpl\ll 1$.
Ultimately,~\cite{deSouza:2023sqp} contains a modified form of equation \eqref{eq:scalar_eom_using_Jordan_frame_trace} that differs from the correct equation by a spurious exponential factor
\begin{equation}
\label{eq:deSouza_Rosenfeld_incorrect_scalar_EOM}  \ddot{\phi}+3H\dot\phi+V'(\phi)=-\frac{g_\nu\beta}{2\pi^2\mpl}{T}_\nu^4\tau\left(\frac{m_\nu}{{T}_\nu}\right)e^{\beta\frac{\phi}{\mpl}}.
\end{equation}
The presence of this extra factor, coupled with extrapolating the model beyond the regime of validity of the EFT results in an expression for the kick $\Delta\phi$ that has a maximum. While this is a critical problem with the results in~\cite{deSouza:2023sqp}, it is not the only issue since, as we will show in the next section, the kick magnitude cannot determine whether a model can or cannot solve the Hubble tension problem.

\subsection{Naturalness of $\nu$EDE}

\begin{figure}[h!]
\centering
\includegraphics[scale=0.8]{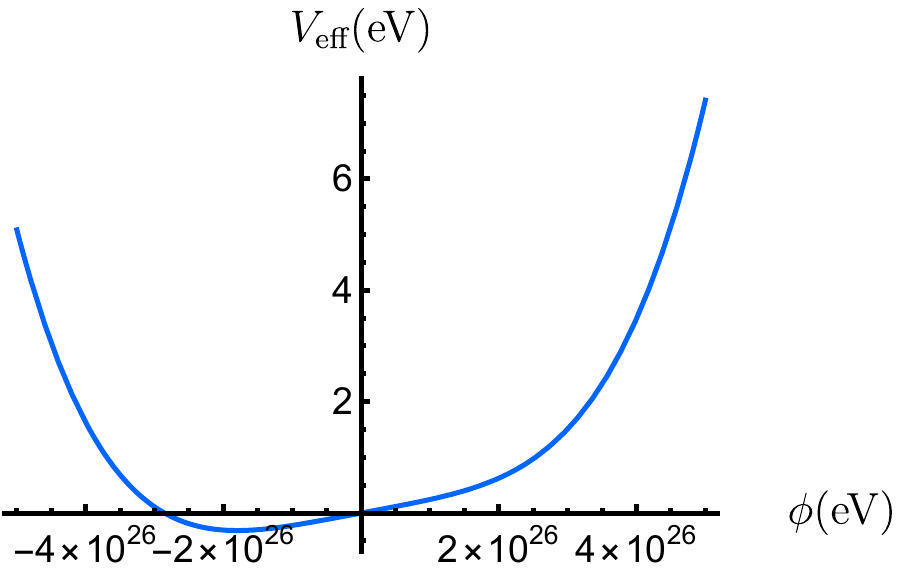} 
\caption{An example of the effective potential in Eq.\eqref{eq:effective_potential} at initial redshift $z_i =5000 $, with parameters  $\lambda = 10^{-98}$, $\beta = 800$, and $m_\nu = 0.3$eV.~It is evident that $\phi_{\text{min}}$ is driven away from $0$.}
\label{fig:Veff}
\end{figure}

We now turn to claims regarding the naturalness and cosmological relevance of $\nu$EDE, beginning with the concept of naturalness.~The argument in~\cite{deSouza:2023sqp} is that, taking $V(\phi)=\lambda\phi^4/4$, the natural initial condition for $\phi$ is $\phi_i=0$, since this corresponds to the minimum of the potential.~However, this fails to account for the fact that the dynamics of the field are governed by an effective potential \cite{Sakstein:2019fmf,CarrilloGonzalez:2020oac}
\begin{equation}
\label{eq:effective_potential}
    V_{\rm eff}(\phi)=V(\phi)-\beta\Theta(\nu)\frac{\phi}{\mpl}, 
\end{equation}
which is minimized at
\begin{equation}
\label{eq:Veff_minimum}
\phi_{\rm min}=\left(\frac{\beta\Theta(\nu)}{\lambda\mpl}\right)^{\frac13}.
\end{equation}
The natural initial condition is thus $\phi_i=\phi_{\rm min}$~\footnote{In the original $\nu$EDE paper \cite{Sakstein:2019fmf} we set $\Theta(\nu)\approx0$ before the kick since the neutrino is ultra-relativistic in the early universe.~The natural initial condition was therefore taken to be $\phi_i=0$, which minimizes $V(\phi)=\lambda\phi^4/4$.~It was later appreciated~\cite{CarrilloGonzalez:2020oac} that the small but non-zero $\Theta(\nu)$ would drive the minimum far from $\phi=0$, making $|\phi_i|=|\phi_{\rm min}|\gg0$ the natural initial condition.}.~Crucially, $\phi_i$  differs significantly from zero and, in fact, a simple estimate shows
\begin{equation}\frac{|\phi_{\rm min}(z)|}{\mpl}\approx10^{-6}\left(\frac{\beta}{800}\right)^{\frac13}\left(\frac{10^{-98}}{\lambda}\right)^\frac13\left(\frac{m_\nu}{0.3\textrm{eV}}\right)^\frac23(1+z)^{\frac23}.
\end{equation}
This can also be seen in Fig.~\ref{fig:Veff}.

We now turn to the concept of a ``cosmologically interesting" phenomenology for $\nu$EDE.~In~\cite{deSouza:2023sqp}, cosmologically interesting models are definied to be those with a kick occurring at redshift $z \in[1585, 6309]$ that injects a fractional energy density $f_{\mathrm{EDE}} \in[7 \%, 13 \%]$.~It is reasonable to use this as a phenomenological criteria for the original EDE model \cite{Poulin:2018cxd}, since in that case the kick is sharp.~However, this is not how the $\nu$EDE model resolves the Hubble tension.~As was pointed out in \cite{CarrilloGonzalez:2020oac}, in $\nu$EDE the kick has a smaller magnitude but lasts longer, and therefore it is necessary to consider the integrated effect rather than solely the maximum magnitude.~Without a full comparison with cosmological datasets, it is therefore insufficient to falsify the model by just looking at the kick.

Furthermore, the conclusion in~\cite{deSouza:2023sqp} that $\nu$EDE is not ``interesting" is drawn because the analysis fails to find any $\phi_i=0$ models that inject the correct amount of EDE at the redshift of MRE.~The same analysis reveals some parameter values that accomplish this with $\phi_i\ne0$ but dismisses these because $\phi_i\ne0$ is not ``natural".~As we have shown above, $\phi_i\ne0$ is in fact the correct initial condition, and there is a large range of initial conditions that are natural.~Thus, by this definition, $\nu$EDE would be cosmologically interesting.~A further claim in~\cite{deSouza:2023sqp} is that those $\nu$EDE models that do inject $\mathcal{O}(10\%)$ EDE around MRE are still uninteresting because it is possible to find uncoupled models with the same $\lambda$ and initial conditions that accomplish a similar timely injection.~The motivation for $\nu$EDE is that it has theoretically-attractive features addressing both the fine-tuning and coincidence problems that its uncoupled counterparts lack.~In this sense, the existence of such counterparts has no bearing on the appeal of $\nu$EDE.

Finally, before concluding, we remark that the analysis in~\cite{deSouza:2023sqp} does not fully explore the theory parameter space.~First, the values of $\beta$ and $\lambda$ investigated constitute only a small region of the viable parameter space uncovered in \cite{CarrilloGonzalez:2020oac}.~Second, while the sum of the neutrino masses is tightly constrained in the base $\Lambda$CDM scenario, it is too restrictive to a priori fix a particular value (as was done in~\cite{deSouza:2023sqp}) in the $\nu$EDE framework, since the constraints can weaken substantially once new physics is introduced \cite{DiValentino:2021imh,Esteban:2022tzx}.~It is therefore reasonable to consider values away from the Planck best-fit, especially since the EDE-neutrino coupling will induce modifications of the Boltzmann hierarchy for neutrinos \cite{Oldengott:2014qra}.~Any numerical exploration of the $\nu$EDE parameter space should account for this.

\section{Conclusions}

In this brief paper, we have addressed the claims made in~\cite{deSouza:2023sqp} and have demonstrated that the conclusions therein do not hold.~One problem is that the equations used to derive a maximum for the kick are not the correct equations relevant for $\nu$EDE.~Another problem is that, regardless of the equations used, once the correct initial conditions are used, claims regarding whether the model is natural or cosmologically interesting are dramatically altered.~Indeed, for this reason an analysis such as that in~\cite{deSouza:2023sqp} cannot exclude the model, no matter what parameter range is explored.~Certainly, neutrino-assisted early dark energy will be constrained by detailed comparisons with cosmological datasets.~This work is underway, and at present $\nu$EDE remains a natural and interesting potential resolution of the Hubble tension.

\section*{Acknowledgements}
The work of QL and MT is supported in part by US Department of Energy (HEP) Award DE-SC0013528.
\bibliography{ref}

\begin{thebibliography}{17}%
\makeatletter
\providecommand \@ifxundefined [1]{%
 \@ifx{#1\undefined}
}%
\providecommand \@ifnum [1]{%
 \ifnum #1\expandafter \@firstoftwo
 \else \expandafter \@secondoftwo
 \fi
}%
\providecommand \@ifx [1]{%
 \ifx #1\expandafter \@firstoftwo
 \else \expandafter \@secondoftwo
 \fi
}%
\providecommand \natexlab [1]{#1}%
\providecommand \enquote  [1]{``#1''}%
\providecommand \bibnamefont  [1]{#1}%
\providecommand \bibfnamefont [1]{#1}%
\providecommand \citenamefont [1]{#1}%
\providecommand \href@noop [0]{\@secondoftwo}%
\providecommand \href [0]{\begingroup \@sanitize@url \@href}%
\providecommand \@href[1]{\@@startlink{#1}\@@href}%
\providecommand \@@href[1]{\endgroup#1\@@endlink}%
\providecommand \@sanitize@url [0]{\catcode `\\12\catcode `\$12\catcode
  `\&12\catcode `\#12\catcode `\^12\catcode `\_12\catcode `\%12\relax}%
\providecommand \@@startlink[1]{}%
\providecommand \@@endlink[0]{}%
\providecommand \url  [0]{\begingroup\@sanitize@url \@url }%
\providecommand \@url [1]{\endgroup\@href {#1}{\urlprefix }}%
\providecommand \urlprefix  [0]{URL }%
\providecommand \Eprint [0]{\href }%
\providecommand \doibase [0]{http://dx.doi.org/}%
\providecommand \selectlanguage [0]{\@gobble}%
\providecommand \bibinfo  [0]{\@secondoftwo}%
\providecommand \bibfield  [0]{\@secondoftwo}%
\providecommand \translation [1]{[#1]}%
\providecommand \BibitemOpen [0]{}%
\providecommand \bibitemStop [0]{}%
\providecommand \bibitemNoStop [0]{.\EOS\space}%
\providecommand \EOS [0]{\spacefactor3000\relax}%
\providecommand \BibitemShut  [1]{\csname bibitem#1\endcsname}%
\let\auto@bib@innerbib\@empty
\bibitem [{\citenamefont {de~Souza}\ and\ \citenamefont
  {Rosenfeld}(2023)}]{deSouza:2023sqp}%
  \BibitemOpen
  \bibfield  {author} {\bibinfo {author} {\bibfnamefont {D.~H.~F.}\
  \bibnamefont {de~Souza}}\ and\ \bibinfo {author} {\bibfnamefont
  {R.}~\bibnamefont {Rosenfeld}},\ }\href@noop {} {\  (\bibinfo {year}
  {2023})},\ \Eprint {http://arxiv.org/abs/2302.04644} {arXiv:2302.04644
  [astro-ph.CO]} \BibitemShut {NoStop}%
\bibitem [{\citenamefont {Verde}\ \emph {et~al.}(2019)\citenamefont {Verde},
  \citenamefont {Treu},\ and\ \citenamefont {Riess}}]{Verde:2019ivm}%
  \BibitemOpen
  \bibfield  {author} {\bibinfo {author} {\bibfnamefont {L.}~\bibnamefont
  {Verde}}, \bibinfo {author} {\bibfnamefont {T.}~\bibnamefont {Treu}}, \ and\
  \bibinfo {author} {\bibfnamefont {A.~G.}\ \bibnamefont {Riess}},\ }\href
  {\doibase 10.1038/s41550-019-0902-0} {\bibfield  {journal} {\bibinfo
  {journal} {Nature Astron.}\ }\textbf {\bibinfo {volume} {3}},\ \bibinfo
  {pages} {891} (\bibinfo {year} {2019})},\ \Eprint
  {http://arxiv.org/abs/1907.10625} {arXiv:1907.10625 [astro-ph.CO]}
  \BibitemShut {NoStop}%
\bibitem [{\citenamefont {Knox}\ and\ \citenamefont
  {Millea}(2020)}]{Knox:2019rjx}%
  \BibitemOpen
  \bibfield  {author} {\bibinfo {author} {\bibfnamefont {L.}~\bibnamefont
  {Knox}}\ and\ \bibinfo {author} {\bibfnamefont {M.}~\bibnamefont {Millea}},\
  }\href {\doibase 10.1103/PhysRevD.101.043533} {\bibfield  {journal} {\bibinfo
   {journal} {Phys. Rev. D}\ }\textbf {\bibinfo {volume} {101}},\ \bibinfo
  {pages} {043533} (\bibinfo {year} {2020})},\ \Eprint
  {http://arxiv.org/abs/1908.03663} {arXiv:1908.03663 [astro-ph.CO]}
  \BibitemShut {NoStop}%
\bibitem [{\citenamefont {Di~Valentino}\ \emph {et~al.}(2021)\citenamefont
  {Di~Valentino}, \citenamefont {Mena}, \citenamefont {Pan}, \citenamefont
  {Visinelli}, \citenamefont {Yang}, \citenamefont {Melchiorri}, \citenamefont
  {Mota}, \citenamefont {Riess},\ and\ \citenamefont
  {Silk}}]{DiValentino:2021izs}%
  \BibitemOpen
  \bibfield  {author} {\bibinfo {author} {\bibfnamefont {E.}~\bibnamefont
  {Di~Valentino}}, \bibinfo {author} {\bibfnamefont {O.}~\bibnamefont {Mena}},
  \bibinfo {author} {\bibfnamefont {S.}~\bibnamefont {Pan}}, \bibinfo {author}
  {\bibfnamefont {L.}~\bibnamefont {Visinelli}}, \bibinfo {author}
  {\bibfnamefont {W.}~\bibnamefont {Yang}}, \bibinfo {author} {\bibfnamefont
  {A.}~\bibnamefont {Melchiorri}}, \bibinfo {author} {\bibfnamefont {D.~F.}\
  \bibnamefont {Mota}}, \bibinfo {author} {\bibfnamefont {A.~G.}\ \bibnamefont
  {Riess}}, \ and\ \bibinfo {author} {\bibfnamefont {J.}~\bibnamefont {Silk}},\
  }\href {\doibase 10.1088/1361-6382/ac086d} {\bibfield  {journal} {\bibinfo
  {journal} {Class. Quant. Grav.}\ }\textbf {\bibinfo {volume} {38}},\ \bibinfo
  {pages} {153001} (\bibinfo {year} {2021})},\ \Eprint
  {http://arxiv.org/abs/2103.01183} {arXiv:2103.01183 [astro-ph.CO]}
  \BibitemShut {NoStop}%
\bibitem [{\citenamefont {Kamionkowski}\ and\ \citenamefont
  {Riess}(2022)}]{Kamionkowski:2022pkx}%
  \BibitemOpen
  \bibfield  {author} {\bibinfo {author} {\bibfnamefont {M.}~\bibnamefont
  {Kamionkowski}}\ and\ \bibinfo {author} {\bibfnamefont {A.~G.}\ \bibnamefont
  {Riess}},\ }\href@noop {} {\  (\bibinfo {year} {2022})},\ \Eprint
  {http://arxiv.org/abs/2211.04492} {arXiv:2211.04492 [astro-ph.CO]}
  \BibitemShut {NoStop}%
\bibitem [{\citenamefont {Riess}\ \emph {et~al.}(2022)\citenamefont {Riess}
  \emph {et~al.}}]{Riess:2021jrx}%
  \BibitemOpen
  \bibfield  {author} {\bibinfo {author} {\bibfnamefont {A.~G.}\ \bibnamefont
  {Riess}} \emph {et~al.},\ }\href {\doibase 10.3847/2041-8213/ac5c5b}
  {\bibfield  {journal} {\bibinfo  {journal} {Astrophys. J. Lett.}\ }\textbf
  {\bibinfo {volume} {934}},\ \bibinfo {pages} {L7} (\bibinfo {year} {2022})},\
  \Eprint {http://arxiv.org/abs/2112.04510} {arXiv:2112.04510 [astro-ph.CO]}
  \BibitemShut {NoStop}%
\bibitem [{\citenamefont {Brout}\ \emph {et~al.}(2022)\citenamefont {Brout}
  \emph {et~al.}}]{Brout:2022vxf}%
  \BibitemOpen
  \bibfield  {author} {\bibinfo {author} {\bibfnamefont {D.}~\bibnamefont
  {Brout}} \emph {et~al.},\ }\href {\doibase 10.3847/1538-4357/ac8e04}
  {\bibfield  {journal} {\bibinfo  {journal} {Astrophys. J.}\ }\textbf
  {\bibinfo {volume} {938}},\ \bibinfo {pages} {110} (\bibinfo {year}
  {2022})},\ \Eprint {http://arxiv.org/abs/2202.04077} {arXiv:2202.04077
  [astro-ph.CO]} \BibitemShut {NoStop}%
\bibitem [{\citenamefont {Sch\"oneberg}\ \emph {et~al.}(2022)\citenamefont
  {Sch\"oneberg}, \citenamefont {Franco~Abell\'an}, \citenamefont
  {P\'erez~S\'anchez}, \citenamefont {Witte}, \citenamefont {Poulin},\ and\
  \citenamefont {Lesgourgues}}]{Schoneberg:2021qvd}%
  \BibitemOpen
  \bibfield  {author} {\bibinfo {author} {\bibfnamefont {N.}~\bibnamefont
  {Sch\"oneberg}}, \bibinfo {author} {\bibfnamefont {G.}~\bibnamefont
  {Franco~Abell\'an}}, \bibinfo {author} {\bibfnamefont {A.}~\bibnamefont
  {P\'erez~S\'anchez}}, \bibinfo {author} {\bibfnamefont {S.~J.}\ \bibnamefont
  {Witte}}, \bibinfo {author} {\bibfnamefont {V.}~\bibnamefont {Poulin}}, \
  and\ \bibinfo {author} {\bibfnamefont {J.}~\bibnamefont {Lesgourgues}},\
  }\href {\doibase 10.1016/j.physrep.2022.07.001} {\bibfield  {journal}
  {\bibinfo  {journal} {Phys. Rept.}\ }\textbf {\bibinfo {volume} {984}},\
  \bibinfo {pages} {1} (\bibinfo {year} {2022})},\ \Eprint
  {http://arxiv.org/abs/2107.10291} {arXiv:2107.10291 [astro-ph.CO]}
  \BibitemShut {NoStop}%
\bibitem [{\citenamefont {Poulin}\ \emph {et~al.}(2019)\citenamefont {Poulin},
  \citenamefont {Smith}, \citenamefont {Karwal},\ and\ \citenamefont
  {Kamionkowski}}]{Poulin:2018cxd}%
  \BibitemOpen
  \bibfield  {author} {\bibinfo {author} {\bibfnamefont {V.}~\bibnamefont
  {Poulin}}, \bibinfo {author} {\bibfnamefont {T.~L.}\ \bibnamefont {Smith}},
  \bibinfo {author} {\bibfnamefont {T.}~\bibnamefont {Karwal}}, \ and\ \bibinfo
  {author} {\bibfnamefont {M.}~\bibnamefont {Kamionkowski}},\ }\href {\doibase
  10.1103/PhysRevLett.122.221301} {\bibfield  {journal} {\bibinfo  {journal}
  {Phys. Rev. Lett.}\ }\textbf {\bibinfo {volume} {122}},\ \bibinfo {pages}
  {221301} (\bibinfo {year} {2019})},\ \Eprint
  {http://arxiv.org/abs/1811.04083} {arXiv:1811.04083 [astro-ph.CO]}
  \BibitemShut {NoStop}%
\bibitem [{\citenamefont {Sakstein}\ and\ \citenamefont
  {Trodden}(2020)}]{Sakstein:2019fmf}%
  \BibitemOpen
  \bibfield  {author} {\bibinfo {author} {\bibfnamefont {J.}~\bibnamefont
  {Sakstein}}\ and\ \bibinfo {author} {\bibfnamefont {M.}~\bibnamefont
  {Trodden}},\ }\href {\doibase 10.1103/PhysRevLett.124.161301} {\bibfield
  {journal} {\bibinfo  {journal} {Phys. Rev. Lett.}\ }\textbf {\bibinfo
  {volume} {124}},\ \bibinfo {pages} {161301} (\bibinfo {year} {2020})},\
  \Eprint {http://arxiv.org/abs/1911.11760} {arXiv:1911.11760 [astro-ph.CO]}
  \BibitemShut {NoStop}%
\bibitem [{\citenamefont {Carrillo~Gonz\'alez}\ \emph
  {et~al.}(2021)\citenamefont {Carrillo~Gonz\'alez}, \citenamefont {Liang},
  \citenamefont {Sakstein},\ and\ \citenamefont
  {Trodden}}]{CarrilloGonzalez:2020oac}%
  \BibitemOpen
  \bibfield  {author} {\bibinfo {author} {\bibfnamefont {M.}~\bibnamefont
  {Carrillo~Gonz\'alez}}, \bibinfo {author} {\bibfnamefont {Q.}~\bibnamefont
  {Liang}}, \bibinfo {author} {\bibfnamefont {J.}~\bibnamefont {Sakstein}}, \
  and\ \bibinfo {author} {\bibfnamefont {M.}~\bibnamefont {Trodden}},\ }\href
  {\doibase 10.1088/1475-7516/2021/04/063} {\bibfield  {journal} {\bibinfo
  {journal} {JCAP}\ }\textbf {\bibinfo {volume} {04}},\ \bibinfo {pages} {063}
  (\bibinfo {year} {2021})},\ \Eprint {http://arxiv.org/abs/2011.09895}
  {arXiv:2011.09895 [astro-ph.CO]} \BibitemShut {NoStop}%
\bibitem [{\citenamefont {Sakstein}(2014)}]{Sakstein:2014jrq}%
  \BibitemOpen
  \bibfield  {author} {\bibinfo {author} {\bibfnamefont {J.}~\bibnamefont
  {Sakstein}},\ }\emph {\bibinfo {title} {{Astrophysical Tests of Modified
  Gravity}}},\ \href {\doibase 10.17863/CAM.16133} {Ph.D. thesis},\ \bibinfo
  {school} {Cambridge U., DAMTP} (\bibinfo {year} {2014}),\ \Eprint
  {http://arxiv.org/abs/1502.04503} {arXiv:1502.04503 [astro-ph.CO]}
  \BibitemShut {NoStop}%
\bibitem [{\citenamefont {Khoury}\ and\ \citenamefont
  {Weltman}(2004)}]{Khoury:2003rn}%
  \BibitemOpen
  \bibfield  {author} {\bibinfo {author} {\bibfnamefont {J.}~\bibnamefont
  {Khoury}}\ and\ \bibinfo {author} {\bibfnamefont {A.}~\bibnamefont
  {Weltman}},\ }\href {\doibase 10.1103/PhysRevD.69.044026} {\bibfield
  {journal} {\bibinfo  {journal} {Phys. Rev. D}\ }\textbf {\bibinfo {volume}
  {69}},\ \bibinfo {pages} {044026} (\bibinfo {year} {2004})},\ \Eprint
  {http://arxiv.org/abs/astro-ph/0309411} {arXiv:astro-ph/0309411} \BibitemShut
  {NoStop}%
\bibitem [{\citenamefont {Brax}\ \emph {et~al.}(2004)\citenamefont {Brax},
  \citenamefont {van~de Bruck}, \citenamefont {Davis}, \citenamefont {Khoury},\
  and\ \citenamefont {Weltman}}]{Brax:2004qh}%
  \BibitemOpen
  \bibfield  {author} {\bibinfo {author} {\bibfnamefont {P.}~\bibnamefont
  {Brax}}, \bibinfo {author} {\bibfnamefont {C.}~\bibnamefont {van~de Bruck}},
  \bibinfo {author} {\bibfnamefont {A.-C.}\ \bibnamefont {Davis}}, \bibinfo
  {author} {\bibfnamefont {J.}~\bibnamefont {Khoury}}, \ and\ \bibinfo {author}
  {\bibfnamefont {A.}~\bibnamefont {Weltman}},\ }\href {\doibase
  10.1103/PhysRevD.70.123518} {\bibfield  {journal} {\bibinfo  {journal} {Phys.
  Rev. D}\ }\textbf {\bibinfo {volume} {70}},\ \bibinfo {pages} {123518}
  (\bibinfo {year} {2004})},\ \Eprint {http://arxiv.org/abs/astro-ph/0408415}
  {arXiv:astro-ph/0408415} \BibitemShut {NoStop}%
\bibitem [{\citenamefont {Di~Valentino}\ and\ \citenamefont
  {Melchiorri}(2022)}]{DiValentino:2021imh}%
  \BibitemOpen
  \bibfield  {author} {\bibinfo {author} {\bibfnamefont {E.}~\bibnamefont
  {Di~Valentino}}\ and\ \bibinfo {author} {\bibfnamefont {A.}~\bibnamefont
  {Melchiorri}},\ }\href {\doibase 10.3847/2041-8213/ac6ef5} {\bibfield
  {journal} {\bibinfo  {journal} {Astrophys. J. Lett.}\ }\textbf {\bibinfo
  {volume} {931}},\ \bibinfo {pages} {L18} (\bibinfo {year} {2022})},\ \Eprint
  {http://arxiv.org/abs/2112.02993} {arXiv:2112.02993 [astro-ph.CO]}
  \BibitemShut {NoStop}%
\bibitem [{\citenamefont {Esteban}\ and\ \citenamefont
  {Salvado}(2022)}]{Esteban:2022tzx}%
  \BibitemOpen
  \bibfield  {author} {\bibinfo {author} {\bibfnamefont {I.}~\bibnamefont
  {Esteban}}\ and\ \bibinfo {author} {\bibfnamefont {J.}~\bibnamefont
  {Salvado}},\ }\href {\doibase 10.22323/1.398.0264} {\bibfield  {journal}
  {\bibinfo  {journal} {PoS}\ }\textbf {\bibinfo {volume} {EPS-HEP2021}},\
  \bibinfo {pages} {264} (\bibinfo {year} {2022})}\BibitemShut {NoStop}%
\bibitem [{\citenamefont {Oldengott}\ \emph {et~al.}(2015)\citenamefont
  {Oldengott}, \citenamefont {Rampf},\ and\ \citenamefont
  {Wong}}]{Oldengott:2014qra}%
  \BibitemOpen
  \bibfield  {author} {\bibinfo {author} {\bibfnamefont {I.~M.}\ \bibnamefont
  {Oldengott}}, \bibinfo {author} {\bibfnamefont {C.}~\bibnamefont {Rampf}}, \
  and\ \bibinfo {author} {\bibfnamefont {Y.~Y.~Y.}\ \bibnamefont {Wong}},\
  }\href {\doibase 10.1088/1475-7516/2015/04/016} {\bibfield  {journal}
  {\bibinfo  {journal} {JCAP}\ }\textbf {\bibinfo {volume} {04}},\ \bibinfo
  {pages} {016} (\bibinfo {year} {2015})},\ \Eprint
  {http://arxiv.org/abs/1409.1577} {arXiv:1409.1577 [astro-ph.CO]} \BibitemShut
  {NoStop}%
\end{thebibliography}%

\end{document}